# High-fidelity ptychographic imaging of highly periodic structures enabled by vortex high harmonic beams

Bin Wang[1*†], Nathan J. Brooks[1*], Peter C. Johnsen[1], Nicholas W. Jenkins[1], Yuka Esashi[1], Iona Binnie[1], Michael Tanksalvala[1], Henry C. Kapteyn[1,2], Margaret M. Murnane[1]

[1]*STROBE Science and Technology Center, JILA, University of Colorado, Boulder, CO 80309, USA*
[2]*KMLabs Inc., 4775 Walnut St. #102, Boulder, CO 80301, USA*
*\*These authors contributed equally*
*†bin.wang-2@colorado.edu*

## Abstract

Ptychographic Coherent Diffractive Imaging enables diffraction-limited imaging of nanoscale structures at extreme ultraviolet and x-ray wavelengths, where high-quality image-forming optics are not available. However, its reliance on a set of diverse diffraction patterns makes it challenging to use ptychography to image highly periodic samples, limiting its application to defect inspection for electronic and photonic devices. Here, we use a vortex high harmonic light beam driven by a laser carrying orbital angular momentum to implement extreme ultraviolet ptychographic imaging of highly periodic samples with high fidelity and reliability. We also demonstrate, for the first time to our knowledge, ptychographic imaging of an isolated, near-diffraction-limited defect in an otherwise periodic sample using vortex high harmonic beams. This enhanced metrology technique can enable high-fidelity imaging and inspection of highly periodic structures for next-generation nano, energy, photonic and quantum devices.

## Introduction

In recent decades, a powerful coherent diffractive imaging (CDI) technique known as ptychography has enabled robust, diffraction-limited, phase-contrast imaging of nanoscale structures [1-5]. Although ptychography has been implemented using a range of illumination wavelengths, its use in the extreme-ultraviolet (EUV) and x-ray regions is particularly attractive for achieving high spatial resolution with inherent elemental and chemical contrast [6-10]. In ptychography, a coherent illumination (the probe) is focused and scanned across an extended sample. A series of far-field diffraction patterns are recorded, while maintaining a large overlap between adjacent scan positions. Iterative phase-retrieval algorithms [11-15] can then be used to robustly and uniquely reconstruct the complex-valued probe field and sample transmission or reflection functions. However, successful reconstruction relies heavily on diversity in the data provided by the lateral scanning of the probe relative to the sample, i.e., interferences at the detector plane mix amplitude and phase, allowing the reconstruction algorithms to unravel both. this means that ptychographic imaging of highly periodic samples with a sufficiently small period is extremely challenging due to the lack of diversity in a series of diffraction patterns, leading to poor convergence of the reconstructed sample images. This significantly limits ptychography's application to a wide variety of nanoscale periodic structures such as photonic crystals [16-17], semiconductor devices [18], and EUV photomasks [19-25]. Consequently, it is critical to fill this characterization gap to aid the advancement of a host of next-generation nano-devices.

High harmonic upconversion of femtosecond lasers can produce bright coherent beams from the EUV to the soft x-ray regions of the spectrum, in a tabletop-scale setup [26-28]. When combined with ptychography, high harmonic generation (HHG) enables phase-sensitive lensless imaging with diffraction-limited nanoscale spatial resolution and excellent elemental specificity [9,15,29-31]. Moreover, because of the quantum-coherent nature of the HHG upconversion process, polarization and phase structure present in the driving laser beam can be transferred to the generated harmonics, provided energy, spin and orbital angular momentum are conserved [32,33]. This makes it possible to create designer short-wavelength structured light for a variety of applications in advanced spectro-microscopies [34,35].

Light beams carrying orbital angular momentum (OAM) have attracted considerable interest for super-resolution imaging [36] and enhanced optical sensing, communication and inspection [37-39]. Recently, by considering conservation of OAM in HHG upconversion, additional routes for controlling the OAM, polarization, as well as the spectral and temporal properties of HHG have been revealed [40-47]. A property particularly interesting for ptychography is the relationship between OAM and the HHG beam divergence/propagation: the spiral phase structure characteristic of OAM-HHG beams forces them to diverge more quickly from the focal point [46]. This means that by using one or more OAM beams to drive the HHG process (referred to as OAM-HHG), one can control the divergence of the emitted HHG probe without changing the focusing optics of the HHG driving laser.

In this article, we demonstrate a solution to a decade-long challenge by showing that high harmonic beams carrying orbital angular momentum can be used to advantage in high-resolution, high-fidelity and fast-convergence ptychographic imaging of highly periodic two-dimensional (2D) grating structures, using the standard extended ptychographic iterative engine (ePIE) algorithm [13]. The key to this technique is that the increased divergence of the OAM-HHG source, combined with the ring-shaped intensity distribution, introduces strong interference fringes between adjacent diffraction orders in the far-field. These encode the non-measurable phase information into the measurable intensity modulation in diffraction fields, significantly enhancing data diversity so that the phase of the diffracted field can be reliably retrieved. We further show that using OAM-HHG beams for illumination provides three significant advantages compared to standard Gaussian-HHG beams, all of which lead to enhanced signal-to-noise-ratio (SNR) for imaging periodic structures: First, due to the conservation of OAM in the HHG process and the resulting strong spiral phase structure in the generated EUV beams, OAM-HHG beams naturally have a significantly increased divergence compared to that of Gaussian-HHG beams. This enhanced illumination NA makes it possible to achieve overlap between different diffraction orders for small pitch periodic samples, beyond what is possible with a Gaussian-HHG probe, and without making any changes to the focusing optics of the ptychographic EUV microscope. Second, the unique ring-shaped OAM beam intensity distribution, which is determined by the strong spiral phase structure in the EUV beams, leads to overlap between different diffraction orders in the high-intensity regions of the beam. And third, OAM-HHG also allows a higher total number of photons to be collected by the detector given a fixed detector dynamic range. Therefore, by leveraging OAM-HHG beams for ptychography, we successfully imaged highly periodic samples with substantially reduced gridding artifacts, and reliably detected defects near the diffraction limit. This new structured-EUV HHG metrology technique can support the advancement of next-generation EUV lithography, nanoelectronics, photonic and quantum devices.

**Methodology**

To date, imaging highly periodic structures has been extremely challenging for ptychography. In a conventional implementation of ptychography using a Gaussian EUV beam to illuminate highly periodic 2D structures (see Fig. 1(a)), the far-field diffraction patterns consist of many isolated diffraction orders, each of which is a copy of the far-field beam profile, and is modulated by an envelope in both amplitude and phase. The zoomed-in green circle in Fig. 1(a) shows this characteristic behavior, with the white circles indicating the edge of each diffraction order. In the resulting ptychographic dataset, diffraction patterns taken at different positions on the highly periodic sample are almost identical to each other. This is because, in contrast to diffraction from non-periodic structures, changes in the far-field diffraction field happen almost entirely in the relative phase between the diffraction orders, but not in the intensity (i.e., diffraction efficiency) of the diffracted orders. The phase information is thus totally lost in this case. Ptychography, as a phase retrieval algorithm, tries to retrieve the phase distributions of diffraction patterns from their intensity measurements. The fact that the phase information is totally lost for highly periodic samples with a sufficiently short period, as opposed to being encoded in the intensity measurements as would be the case for non-periodic structures, makes it extremely challenging to achieve successful ptychographic imaging of such highly periodic structures. As expected, the reconstruction fails for ptychography using a Gaussian-HHG beam, as shown in Fig. 1(b), in which the amplitude and phase of the reconstruction are plotted in brightness and hue, respectively. This phase problem can also be understood through the convolution theorem, as discussed in detail in Supplementary Section 1.

In 1969, Hoppe proposed to achieve electron diffraction imaging of periodic atomic lattices by encoding the non-measurable phase information into the measurable intensity modulation in diffraction patterns, through interference between neighboring diffraction orders [48]. As schematically shown in Fig. 1(c), OAM-HHG beams are able to achieve overlap and interference between neighboring diffraction orders due to their intrinsically larger beam divergence and ring-shaped intensity distribution. The zoomed-in blue circle in Fig. 1(c) shows the interference fringes, with the yellow circles indicating the edge of each diffraction order. As one scans the probe relative to the periodic structures, the relative phase of each diffraction order changes accordingly, which then causes the interference fringes to shift. In other words, the phase information in the diffraction patterns is now encoded in the intensity measurements through interference. These interference fringes contain the missing phase information, and increase the diversity in diffraction patterns, thereby enabling robust and reliable ptychographic reconstructions (see Supplementary Section 2). Figure 1(d) shows a high-fidelity ptychographic reconstruction of a 2D periodic structure under an OAM-HHG illumination.

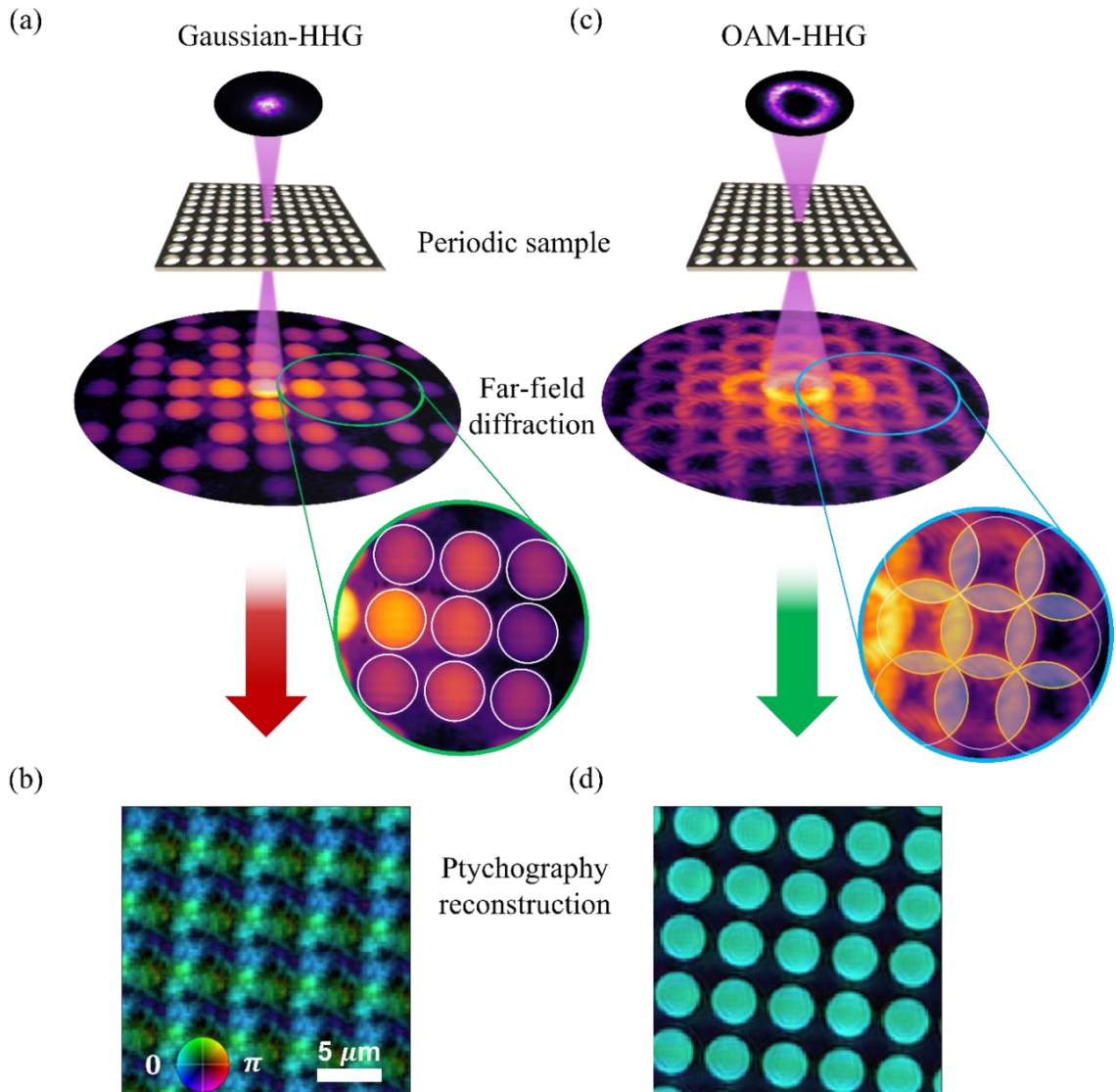

**Figure 1. Robust and reliable ptychographic imaging of highly periodic structures.** (**a**) Schematic showing HHG ptychographic imaging of a periodic structure using conventional Gaussian-HHG illumination. The resulting diffraction orders are isolated (see zoomed-in green circle), where the white circles indicate the edges of each diffraction order. This leads to a complete loss of the relative phase information between the orders in the far-field diffraction, which subsequently leads to the failure of the ptychographic reconstruction in (**b**). (**c**) OAM-HHG illumination intrinsically has a larger source divergence and a ring-shaped intensity profile, to support overlap and interference between diffraction orders (see zoomed-in blue circle), in which the yellow circles indicate the edges of each diffraction order. This interference converts the relative phase between the diffraction orders into measurable intensity modulation, enabling fast and robust ptychographic reconstruction of the 2D periodic structure in (**d**). In (**b**, **d**), the complex-valued amplitude and phase are plotted as brightness and hue, respectively.

The required NA for high fidelity imaging of periodic samples can be understood as follows. When a periodic structure is illuminated by a focused coherent beam, the angular separation between two neighboring diffraction orders is given by $\Delta\theta = \lambda/\Lambda$, where $\lambda$ is the illumination wavelength, and $\Lambda$ is the period of the structure. The illumination NA for the microscope is defined as the half-cone angle of the focusing beam. Geometrically, for fixed $\lambda$ and $\Lambda$, there exists a critical value for illumination NA:

$$NA_c = \frac{1}{2}\Delta\theta = \frac{\lambda}{2\Lambda}. \tag{1}$$

Only for illumination NAs larger than $NA_c$ will neighboring diffraction orders have sufficiently large footprints on the detector to overlap with each other and generate interference fringes, thus enabling successful ptychographic reconstructions.

## Experimental configuration

We designed and built an EUV ptychographic microscope in a transmission geometry, as shown in Fig. 2. The driving laser for the HHG process is a frequency-doubled Ti:sapphire laser amplifier system ($\lambda\sim 395$ nm), with an intrinsic near-Gaussian mode (vortex charge of $\ell = 0$), that can be converted to an OAM beam of vortex charge $\ell = 1$ using a spiral phase plate. The 7th harmonic of the driving laser ($\lambda\sim 56$ nm) exhibits either a Gaussian mode or an OAM of vortex charge $\ell = 7$ depending on whether a spiral phase plate is used. The EUV beam is then focused by a double-toroid focusing system onto the periodic samples, with a spot size of ~13 × 18 μm ($1/e^2$ intensity) for Gaussian-mode HHG, or ~27 × 32 μm (donut intensity peak-to-peak) for OAM-HHG. The reconstructed Gaussian- and OAM-HHG beam profiles are shown in a complex representation in Fig. S4, with the beam amplitude and phase indicated by brightness and hue. The test samples are three Quantifoil holey carbon films (~20 nm thick) which have various hole sizes and shapes arranged in a periodic rectangular grid. The three Quantifoil holey carbon films have a pitch of 9 μm, 4.5 μm and 3 μm, respectively. These Quantifoil holey carbon films are mounted on standard Ted Pella Ø3mm Cu 200 mesh TEM grids. (See the Methods section for more information.)

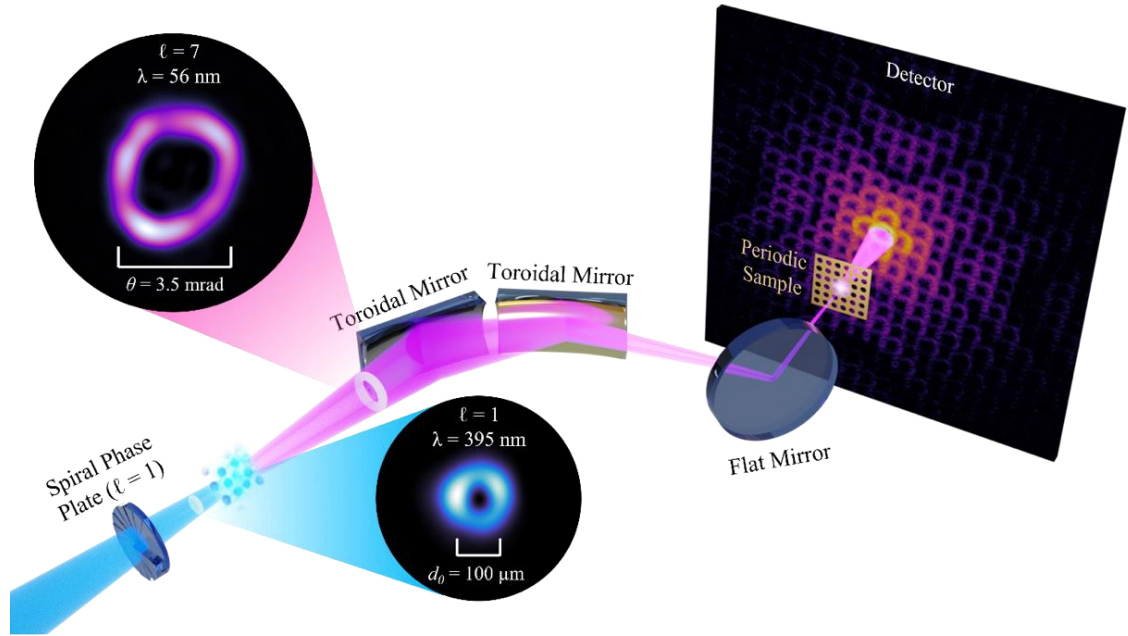

**Figure 2. EUV ptychographic microscope using OAM-HHG EUV beams for imaging highly periodic structures.** A spiral phase plate ($\ell$ = 1 at 395 nm wavelength) converts the driving laser at 395 nm wavelength to an OAM beam, which is focused into a semi-infinite gas cell to produce a nearly monochromatic 7th harmonic beam with a wavelength of 56 nm and an OAM charge of $\ell$ = 7. A double-toroidal mirror focusing system focuses this OAM-HHG beam onto a 2D periodic sample, and an EUV-CCD camera is used to record the far-field diffraction patterns.

During the ptychography scans, the test samples are translated in the plane perpendicular to the beam path in $7 \times 7$ rectangular grids (49 scan positions) with nominally 3.3 µm distance between adjacent scan positions. A random offset of $\pm 20\%$ of the scan step size was added to each scan position to avoid gridding artifacts in the reconstructions [49]. The far-field diffraction patterns are recorded by an EUV-CCD detector (Andor iKon-L, $2048 \times 2048$, 13.5 µm pixel size) positioned 50 mm after the sample. To obtain the best ptychographic reconstructions possible for each illumination case, we carefully pre-characterized each probe function in the sample plane by taking ptychographic scans on a non-periodic sample and reconstructing both the sample and the probe functions through blind deconvolution. These pre-characterized probe functions were used as initial guesses in the ptychographic reconstructions of highly periodic structures. Other than using the pre-characterized probe as the initial guess, we used the standard ePIE algorithm [13] for all reconstructions in this study, without the need for additional constraints such as modulus enforced probe [15] or total variation regularization [25,50].

## Results

*OAM-HHG enables robust and reliable ptychographic imaging of periodic structures*

We experimentally demonstrate that OAM-HHG beams enable robust and reliable ptychographic imaging of highly periodic structures because of three intrinsic advantages compared to Gaussian-HHG illumination. First, due to the conservation of OAM in the HHG process and the resulting strong spiral phase structure in the generated EUV beams, OAM-HHG beams naturally exhibit a significantly increased divergence (i.e., increased illumination NA for the microscope given the same focusing optics) compared with Gaussian-HHG beams. This enhanced illumination NA allows us to achieve overlap between diffraction orders for smaller pitch periodic samples beyond what is possible with a Gaussian-mode probe, without making any changes to the EUV microscope end-station. Second, the characteristic ring-shaped intensity distribution of OAM-HHG beams ensures that the majority of photons fall in the overlap area (in contrast to the Gaussian-HHG beams, for which the overlap between diffraction orders occurs at the tails of the intensity distributions), which increases the SNR for the interference fringes. Third, the ring-shaped intensity distribution of OAM-HHG beams allows one to collect a higher total number of photons by the detector given a fixed dynamic range, which also leads to higher SNR without the need for high dynamic range (HDR).

We performed ptychographic imaging on three highly periodic structures with 9 µm, 4.5 µm and 3 µm pitches using Gaussian- and OAM-HHG beams at a wavelength of 56 nm. Example diffraction patterns and reconstructed images from each ptychography scan can be found in Fig. 3. Furthermore, each ptychography scan collected 49 far-field diffraction intensity patterns, as shown in a log scale in Supplementary Video 1. Note that while there is a clear change in the diffraction patterns from frame to frame for the OAM-HHG case, particularly in the interference fringes between the adjacent diffraction orders, the diffraction patterns in the Gaussian-HHG case do not change very much for small period samples. All ptychography datasets were taken without making any changes to the EUV microscope — the difference in divergence between Gaussian- and OAM-HHG beams is intrinsic to the HHG upconversion process, which conserves energy and OAM.

For the 9 µm pitch sample, the small diffraction angle means that successive orders largely overlap even for the Gaussian-HHG beam, as shown in Fig. 1(a). This results in a reasonably good image, apart from some gridding artifacts as shown in Fig. 1(d). In comparison, the ptychography scan using OAM-HHG illumination sees more overlap resulting in improved SNR in the interference fringes and a much higher-fidelity image with greatly reduced gridding artifacts, as shown in Fig. 1(g,j).

For the smaller 4.5 µm pitch sample, the diffraction orders are further apart, causing the Gaussian-HHG beams to lose most of the interference in the diffraction patterns, as shown in Fig. 3(b). The low SNR in the interference fringes results in reduced quality image reconstruction, as shown in Fig. 3(e). However, due to their higher intrinsic divergence and the unique ring-shaped intensity distribution, OAM-HHG maintains a large area of overlap between neighboring diffraction orders with more photons, as shown in Fig. 3(h). This results in higher-quality images of the periodic structure with a 4.5 µm period, as shown in Fig. 3(k). Thus, simply by inserting a spiral phase plate to convert the driving laser to an OAM beam, while keeping everything else in the microscope the same, a greatly improved reconstruction quality is obtained.

Lastly, for the smallest 3 μm period sample, Gaussian-HHG illumination totally fails due to the lack of interference between diffraction orders, as shown in Fig. 3(c,f). In this case, OAM beams can reconstruct a reasonable image, although the quality of the unit cell is degraded, as shown in Fig. 3(i,l).

We also evaluated the quality of these ptychographic reconstructions using complex histogram analysis and the results can be found in Supplementary Section 3. We note that all reconstructions in Fig. 3 have the correct sample periodicity because this information is directly available from the measured intensity of the diffracted fields — the success or failure of the reconstructions of the unit cells depends on whether the relative phase between the diffraction orders can be successfully retrieved or not. The ptychography reconstructed images in Fig. 3(d–f, j–l) are complex-valued and are shown in a complex representation where the amplitude and phase information are represented by the brightness and hue, respectively. The color wheel is shown in the bottom left corner of each panel.

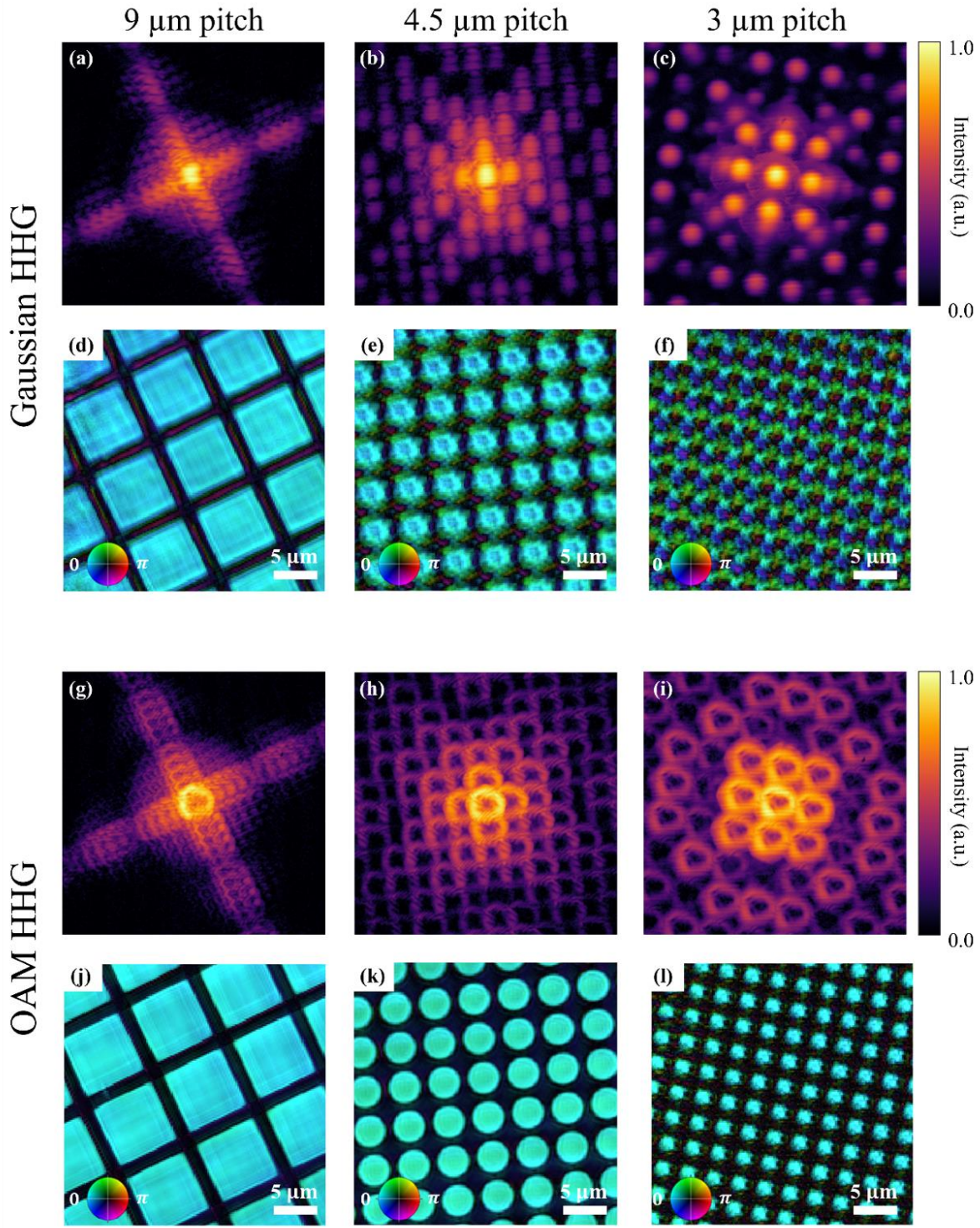

**Figure 3. High divergence OAM-HHG beams are able to produce higher-quality ptychography images of periodic structures than low divergence Gaussian-HHG beams.** Three test samples with different periods and shapes, i.e., 9 µm period with square holes, 4.5 µm period with circular holes and 3 µm period with circular holes, are investigated. For Gaussian-HHG beams, example diffraction patterns from the three test samples are shown in (a–c) and the corresponding ptychography reconstructed images are shown in (d–f). The example diffraction patterns and ptychography images from OAM-HHG beams are shown in (g–i) and (j–l). The complex-valued image in (**d–f, j–l**) are plotted in a complex representation where amplitude and phase are shown in brightness and hue, respectively.

*OAM-HHG beams reveal nanoscale defects in otherwise periodic samples*

A major motivation for imaging periodic structures is to reliably detect and pinpoint small areas where the periodicity is broken, i.e., to locate defects. However, when the diffraction orders are insufficiently overlapped, the artifacts in the ptychographic reconstructions make it difficult or even impossible to locate defects. In contrast, the increase in reconstruction quality enabled by OAM-HHG beams, especially the suppression of periodic artifacts in the reconstructions (inherent to ptychographic imaging of periodic structures), enables reliable location of nanoscale defects in otherwise highly periodic structures. This can potentially find its application in metrology for micro- and nano-fabrication and manufacturing, including in advanced metrologies in support of EUV lithography.

In the 9 μm pitch sample, a damaged carbon bar (~300 nm wide) can be seen in the scanning electron microscopy (SEM) image in Fig. 4(e), indicated by the red arrow. We first performed ptychographic imaging of the corresponding area of the sample using an OAM-HHG beam. During data acquisition at each scan position, we acquired two diffraction patterns with exposure times of 0.1 and 1 second, and combined them to form a composite high dynamic range (HDR) image to increase SNR. The reconstructed image of the transmitted amplitude is shown in Fig. 4(a), in which the defect is clearly resolved and is indicated by the red arrow. Given that the pixel size in the ptychography reconstruction images is 200 nm, the fact that our EUV microscope using OAM-HHG illuminations can clearly image a defect with size of about 300 nm (i.e., 1.5× the pixel size in the reconstruction images) makes it very promising to detect or image smaller defects down to 10's of nanometers using shorter EUV wavelengths and increased imaging NA.

Next, a similar experiment is performed using a Gaussian-HHG beam under the same conditions, resulting in the same approximate maximum detector count in the diffraction patterns as for the OAM case. The reconstructed image of the transmitted amplitude is shown in Fig. 4(b), where reconstruction artifacts heavily corrupt the image details and render the defect unidentifiable. Furthermore, due to the different intensity distributions of the Gaussian- and OAM-HHG beams, even though the two datasets have the same maximum detector count, the OAM dataset has 3 times more total detector counts than the Gaussian one.

To confirm that the difference in reconstructed image quality is not simply due to this different in the total number of photons collected, but is due to how those photons are distributed in the diffraction plane (i.e., in the area of overlap between diffraction orders), we performed a third experiment using the Gaussian-HHG beam and triple HDR exposure (0.1-, 1- and 3-second exposure time), which leads to longer data acquisition time by a factor of 1.67, to have approximately equal total counts in the combined diffraction data compared to the OAM-HHG case. The resulting amplitude image is shown in Fig. 4(c). There is significant improvement over the reconstructed image from Gaussian-HHG beams with double HDR exposure, but the reconstruction artifacts still make it difficult to identify the nanoscale defect. We further quantitatively analyzed the SNR of the defect in these three reconstruction amplitude images using the transmission profiles of the thin carbon bar in the boxes in Fig. 4(a-c). These transmission profiles are obtained by averaging the transmission images in the vertical direction, and are plotted along the horizontal direction, as shown in Fig. 4(d). The SNRs of the defect in Figs. 4(a–c) are calculated (see Methods) and summarized in Table 1. The SNR for the defect image from OAM-HHG illumination is improved by a factor of >135 compared to that from Gaussian-HHG illumination with equal exposure time. Furthermore, we evaluated the quality of these ptychographic reconstructions using complex histogram analysis and verified that OAM-HHG illuminations result in higher fidelity images, as discussed in detail in Supplementary Section 3. It is worth emphasizing that when taking the SEM image in Fig. 4(e), the high-energy electron beam at 300 keV severely damaged the thin carbon bar, causing shrinkage and the appearance of the bright areas on the top and bottom edges. In contrast, the EUV HHG beam can non-destructively image both the periodic sample and the defect.

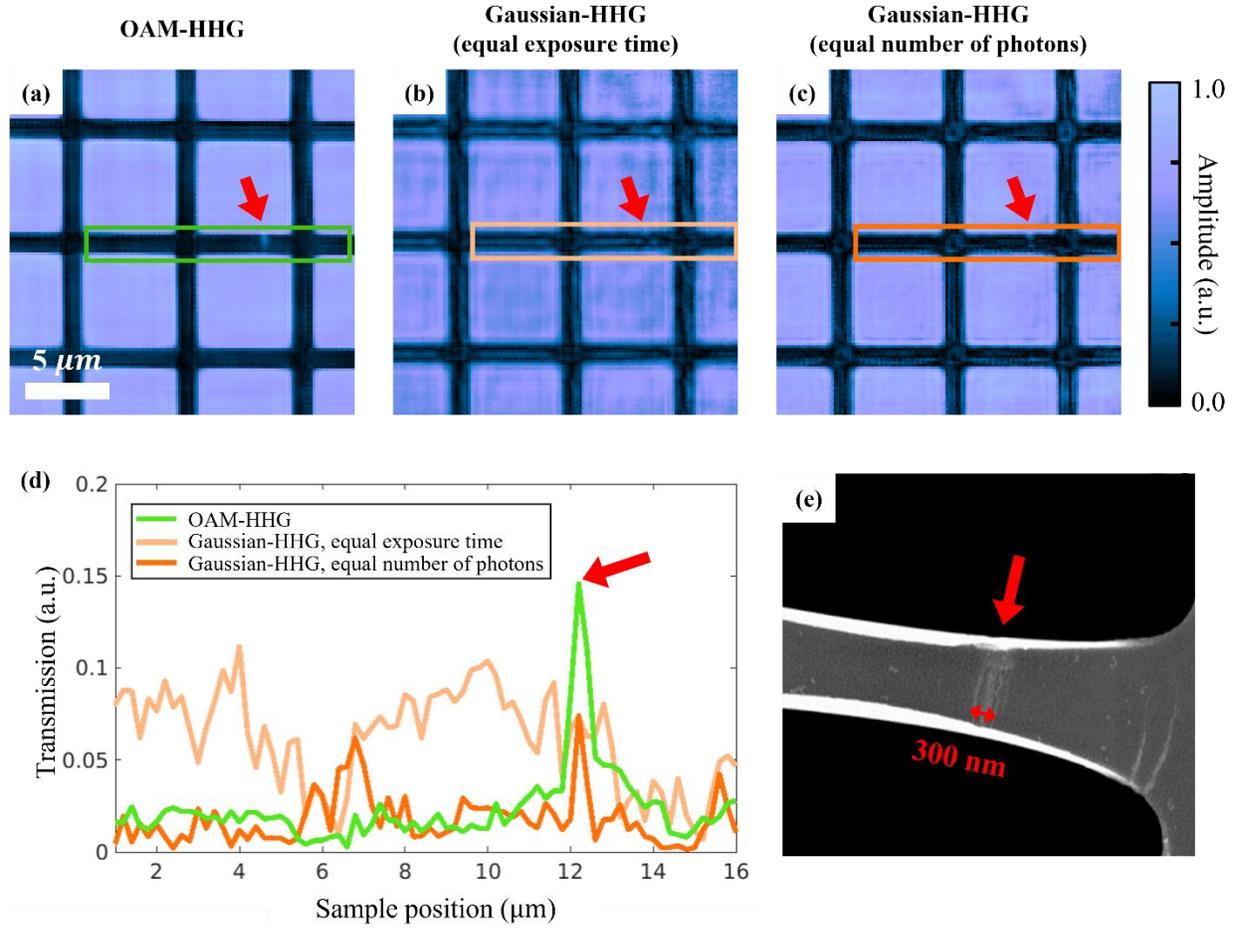

**Figure 4. Enhanced sensitivity to nanoscale defects in periodic structures using OAM-HHG beams.** (**a–c**) Amplitude images of ptychographic reconstruction of a 2D square periodic structure (9 µm period, with a nanoscale defect of ~300 nm in size) under various conditions: (**a**) OAM-HHG beams with double HDR (0.1- and 1- second exposure times), (**b**) Gaussian-HHG beams with equal exposure time as the OAM-HHG case using double HDR (0.1- and 1- second exposure times), and (**c**) Gaussian-HHG beams with roughly equal number of photons as the OAM-HHG case using triple HDR (0.1-, 1- and 3-second exposure times). The red arrows indicate the nano-defect in the thin carbon bar. (**d**) Ptychography reconstructed transmission profile of the thin carbon bar containing a nano-defect (indicated by the red arrow) in the boxes in (**a**–**c**). The transmission profiles are averaged in the vertical direction. The red arrow indicates the nano-defect. (**e**) An SEM image of the same sample area shows a 300-nm-wide defect. Bright areas on the top and bottom edges are due to sample damage from the high energy electron (300 keV) beams.

**Table 1. Signal-to-noise ratio analysis for the three ptychographic reconstructions shown in Fig. 4(a–c).**

| Ptychographic reconstructions | *signal* | *background* | *noise* | *SNR* | Improvement factor |
|---|---|---|---|---|---|
| **OAM-HHG in Fig. 4(a)** | 1.46e-1 | 1.52e-2 | 5.62e-3 | 23.25 | **135.8** |
| **Gaussian-HHG, equal exposure time in Fig. 4(b)** | 7.34e-2 | 6.94e-2 | 2.35e-2 | 0.17 | Benchmark |
| **Gaussian-HHG, equal number of photons in Fig. 4(c)** | 7.40e-2 | 1.72e-2 | 1.31e-2 | 4.34 | 25.53 |

## Conclusion

In conclusion, we demonstrated that by incorporating illumination engineering via OAM-HHG beams into an EUV ptychography microscope, we can address the long-standing challenge of high-fidelity coherent diffractive imaging of periodic structures. The intrinsic large divergence and ring-shaped intensity distribution of OAM-HHG beams leads to the formation of higher SNR interference fringes in the diffraction patterns — thus enabling faster and higher fidelity image reconstructions using the basic ePIE algorithm, without extra algorithmic effort. Furthermore, the improvement in image fidelity allowed sensitive detection of a 300 nm wide defect, which is 1.5× the pixel size of the reconstructed images, in an otherwise periodic thin carbon mesh with 9 μm period.

Ptychographic imaging of highly periodic structures has been widely recognized to be challenging, which has precluded its application in critical science and technology fields such as semiconductor metrology and EUV photomask inspection. Future studies can employ coherent EUV and X-ray vortex beams to enable nanometer- or even sub-nanometer-scale spatial resolution in a broad range of next-generation nanoelectronics, photonics and quantum devices. A particularly interesting direction would be to use coherent EUV light at a wavelength of 13.5 nm for actinic imaging and inspection of EUV photomasks [19-25]. Finally, this work can provide inspiration for the electron ptychography community (e.g., cryo-EM and 4D-STEM), where recent work has explored the potential benefits of engineered vortex electron beams for enhanced imaging fidelity and lower dose [51,52].

## Methods

*Experimental setup*

A Ti:sapphire amplifier system (KMLabs Wyvern HE) with a $\lambda = 790$ nm central wavelength, 45 fs pulse duration, 8 mJ pulse energy, and 1 kHz repetition rate was used for this demonstration. Part of the laser output is used for second harmonic generation (SHG) in a $\beta$-barium borate crystal ($\beta$-BBO), yielding a frequency doubled beam at 395 nm central wavelength for driving the HHG process. This SHG beam is focused into a semi-infinite gas cell, which consists of a Brewster-cut entrance window, a 20 cm length filled with 50 torr of argon gas, and a copper gasket placed in the focal plane of the driving laser where a coherent HHG beam is generated [53,54]. The driving laser at 395 nm central wavelength is separated from the high-harmonic beam by using a 200 nm aluminum filter. This filter also blocks any harmonics with $\lambda > 77$ nm, while harmonics with $\lambda < 39$ nm exceed the HHG cutoff energy, and so are not generated. Furthermore, due to the centrosymmetry of the medium, only odd-numbered harmonic orders are generated. The resulting EUV beam after the aluminum filter thus consists of narrow peaks at the 7$^{th}$ ($\lambda = 56$ nm) and 9$^{th}$ ($\lambda = 44$ nm) harmonics. The intensity ratio of the two harmonics in our experimental setup is estimated to be $I_{\lambda=56nm}/I_{\lambda=44nm} \sim 30:1$, which can be well-approximated as a monochromatic illumination suitable for ptychographic imaging. For generating HHG beams with a Gaussian spatial profile, we used an SHG beam with pulse energy of ~500 μJ. For generating HHG beams carrying OAM, we increased the pulse energy of the SHG beam to ~1.5 mJ, and inserted a spiral phase plate (Holo-Or, VL-214-395-Y-A, OAM charge number $\ell = 1$ at 395 nm wavelength) right after the focusing optics into the semi-infinite gas cell to generate a driving beam with OAM charge number $\ell = 1$, and $\lambda = 395$ nm. The increased pulse energy is necessary in order to make the peak intensity (located at a central point for the Gaussian beam, but distributed in a ring for the OAM beam) equal for the two cases, thus matching HHG cutoff energies and conversion efficiency. Due to the conservation of OAM in HHG, the resulting quasi-monochromatic 7$^{th}$ harmonic beam ($\lambda = 56$ nm) carries an OAM charge number of $\ell = 7$.

The HHG beam at 56 nm wavelength is focused sequentially by two toroidal mirrors (1: B$_4$C-coated, $f_{eff} = 27$ cm, $\theta = 15°$; 2: Au-coated, $f_{eff} = 50$ cm, $\theta = 10°$) in a Wolter configuration to create an imaging system with higher NA ($f_{eff} = 17$ cm) while managing coma aberration [55]. The resulting focusing beam is redirected towards the sample at normal incidence using a glancing incidence mirror (B$_4$C coating, fused silica substrate, 3° incidence angle from grazing, nominal reflectivity 95%). The testing samples are three Quantifoil holey carbon films which have various hole sizes and shapes arranged in a rectangular grid, and are mounted on standard Ted Pella Ø3mm Cu TEM grids with 200 mesh (125 um pitch, 90 um hole width and 35 um bar width). More specifically, the three Quantifoil holey carbon films have a pitch of 9 μm (7 μm square hole and 2 μm bar, product number 656-200-CU), 4.5 μm (3.5 μm diameter circular holes and 1 μm separation, product number 660-200-CU) and 3 μm (2 um diameter circular holes and 1 um separation, product number 661-200-CU), respectively. The samples are positioned close to the beam focus, and are mounted on a precision translation stage ensemble (SmarAct XYZ-SLC17:30). They are translated in the plane perpendicular to the beam path to perform ptychographic scans in 7 × 7 rectangular grids (49 positions) with nominally 3.3 μm between adjacent positions. A random offset of ±20% of the scan step size was added to each scan position to avoid artifacts originating from the scan grid itself. The far-field diffraction patterns are recorded by an EUV-CCD detector (Andor iKon-L, 2048 × 2048, 13.5 μm pixel size) positioned 50 mm after the sample. In order to obtain the best ptychographic reconstructions possible for each illumination case, we carefully characterized each probe function in the sample planes by taking ptychographic scans on a non-periodic sample and reconstructing both the sample and the probe function through blind deconvolution. The reconstructed probe functions were used in the ptychographic reconstructions of highly periodic samples as initial guesses.

*Ptychographic data processing and image reconstructions*

The diffraction patterns were recorded by an EUV-CCD detector with 2048 × 2048 pixels and 13.5 μm detector pixel size. During data processing, we cropped them to 1024 × 1024 because very few photons were detected outside this area. The resulting pixel size of the reconstructed images is

$$dr = \frac{\lambda \cdot z}{N \cdot dx} \approx 200 \text{ nm}, \tag{2}$$

where $\lambda$ = 56 nm is the wavelength, $z$ = 50 cm is the distance from the sample to the CCD detector, $N$ = 1024 is the number of pixels in one direction and $dx$ = 13.5 μm is the detector pixel size.

The ptychographic reconstructions were performed in two steps using only the ePIE algorithm [13]. In the first step, the complex-valued probe functions (both Gaussian- and OAM-HHG beams) were characterized by performing ptychography on a non-periodic sample and using the ePIE algorithm for reconstruction. In the second step, the pre-characterized probe functions were used as initial guesses for reconstructing the periodic structures. For the first 100 iterations, only the sample images were updated while the probe functions were kept fixed. Then, both the sample images and probe functions were updated by the ePIE algorithm for another 900 iterations. The total number of iterations for ptychographic reconstructions of periodic structures was 1000. This procedure was kept constant for all Gaussian-HHG and OAM-HHG reconstructions in this paper. We also want to emphasize the fast convergence speed of our technique compared to that in the work by Gardner et al. [15], which takes more than 10,000 iterations.

*SNR analysis of imaging of the nano-defect*

The SNR of the defect detection in Table 1 is calculated as follows: We start from the three curves in Fig. 4(d). For each curve, corresponding to an experimental condition shown in the legend, the *signal* level is the transmission value in the defect, the *background* level and *noise* level are calculated as the average and the standard deviation, respectively, of the transmission values excluding the defect. The *SNR* is then calculated using the following formula:

$$SNR = \frac{signal - background}{noise}. \tag{3}$$

*Data availability*

The data that supports the plots and other findings within this paper are available from the corresponding author upon reasonable request.

## Acknowledgements

This research was primarily supported by STROBE: a National Science Foundation (NSF) Science and Technology Center (STC) under award DMR-1548924 for the setup and new illumination engineering and algorithms, and also by a DARPA STTR grant 140D0419C0094 for imaging periodic samples. A Moore Foundation Grant No. 10784 supported the low-dose imaging research. The authors thank Guan Gui, Drew Morrill, Yunzhe Shao, Chen-Ting Liao, Emma Cating-Subramanian for comments on the text.

## Author contributions

B. W., N. J. B., M. M. M. and H. C. K. conceived the experiment. B. W., N. J. B. and P. J. built and maintained the EUV source. B. W. and N. J. B. collected the data sets and performed the reconstructions and data analysis. N. J., Y. E. and B. W. performed the SEM imaging of the test samples. M.T., Y.E. and N.J. advised on the phase retrieval algorithms and setup, while I.B. helped to develop the laser setup. All authors designed aspects of the experiment, performed the research and wrote the paper.

## Competing financial interests

# Supplementary information:

## S1. Convolution theorem perspective on ptychographic imaging of highly periodic structures

In ptychography, the far-field diffraction fields are approximated as the Fourier transform of the product of the complex probe and object functions, $p(x,y)$ and $o(x,y)$, i.e.,

$$\Psi(u,v) = \mathcal{F}[p(x,y) \times o(x,y)], \tag{S1}$$

where $\mathcal{F}$ is the Fourier transform operation, $(x,y)$ are the real space coordinates and $(u,v)$ are the reciprocal space coordinates. According to the convolution theorem, this can also be represented as a convolution of the Fourier transform of the probe function, $P(u,v) = \mathcal{F}[p(x,y)]$, and that of the object function, $O(u,v) = \mathcal{F}[o(x,y)]$, i.e.,

$$\Psi(u,v) = P(u,v) * O(u,v), \tag{S2}$$

where $*$ is the convolution operation. Often, $P(u,v)$, which is the complex beam in the detector plane when no sample is in the way, has an edge resulting from apertures in the system, as indicated by the circle in the close-ups in Fig. 1(a,c). In the case of 2D highly periodic structures, $O(u,v)$ consists of a 2D comb of $\delta$ functions (diffraction peaks) arranged in a 2D periodic grid, the amplitudes and phases of which are modulated by the Fourier transform of the unit cell of the periodic structure. This is a sparse function in the reciprocal space. The convolution operation in Eq. (S2) puts a copy of $P(u,v)$ at the location of each $\delta$ function with modulated amplitude and phase.

In cases where $P(u,v)$ is small in size such that all diffraction orders are isolated, the modulated phase of each diffraction order is totally lost when we collect intensity measurements, thus causing ptychography to fail. However, in cases where $P(u,v)$ is sufficiently large in size, the interference fringes in the overlapped regions between neighboring diffraction orders encoded the relative phase of each diffraction orders into the intensity modulations that are directly measurable with the EUV-CCD camera, thus enabling fast and robust ptychographic reconstructions of the highly periodic structures.

## S2. A phase-change-like behavior in ptychography demonstrated by Gaussian-HHG illuminations with controlled divergence

Since the key to successfully achieving ptychographic imaging of highly periodic structures is to obtain overlap and interference between neighboring diffraction orders, an abrupt, phase-change-like behavior in reconstruction quality is expected as one smoothly changes the illumination NA. We experimentally demonstrated this behavior in ptychographic imaging of highly periodic structures, as shown in Fig. S1, using Gaussian-HHG illuminations with a controlled illumination NA. This is achieved by installing an in-vacuum iris ~0.5 m after the semi-infinite gas cell, which allows direct control of the divergence of the HHG beams, and thus of the illumination NA on the sample and the overlap between neighboring diffraction orders given the same focusing optics.

We performed four ptychography scans on the same 2D square periodic structure with a 9 μm period under various illumination NAs controlled by the in-vacuum iris. Fig. S1(a–d) shows example diffraction patterns from each scan from small illumination NA in (a) to large illumination NA in (d). The close-ups in the blue circles show the effect of illumination NA on the resulting diffraction patterns. We then reconstructed these four datasets using the standard ePIE algorithm [13] and the corresponding results are shown in Fig. S1(e–h). It is clear that for ptychography scans where diffraction orders are isolated, the periodic structure cannot be reliably reconstructed due to the lost phase information, as shown in Fig. S1(e–f), while for ptychography scans where the illumination NA is large enough to support overlap between diffraction orders, the ePIE algorithm can quickly and reliably reconstruct the periodic structures.

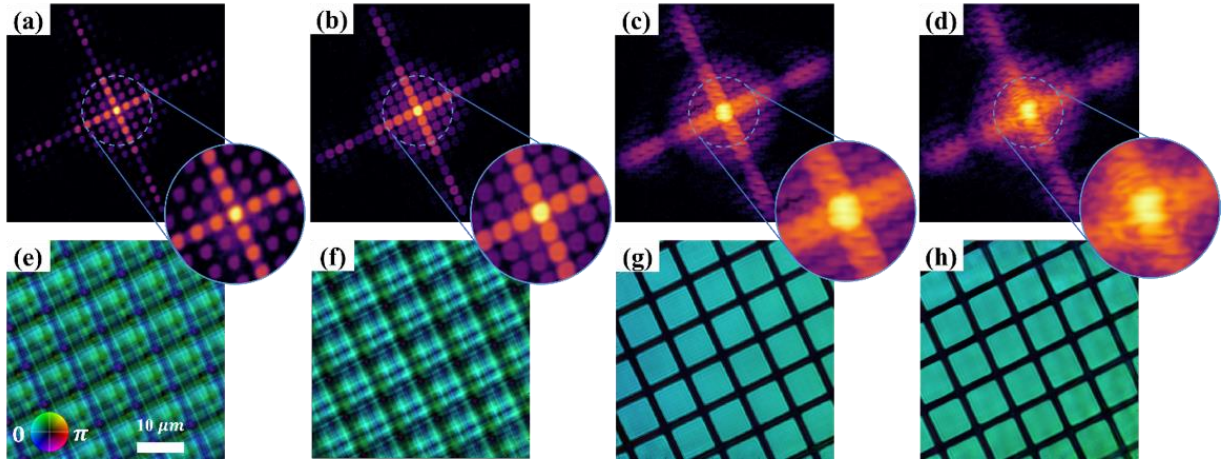

**Figure S1.** Experimental demonstrations of a phase-change-like behavior in ptychographic imaging of 2D square periodic structures with 9 um period using Gaussian-HHG beams with controlled divergence. (**a–d**) Example diffractions from a 2D square periodic structure using Gaussian-HHG beams with various divergences. The inserts show close-ups of the center of the diffraction patterns. (**e–h**) The corresponding ptychographic reconstructions of the 2D square periodic structure under various illumination conditions. The reconstructions are successful only when the diffraction orders have overlap, showing a phase-change-like behavior.

## S3. Image quality assessment using complex histogram analysis

We use a complex histogram analysis to evaluate the quality of the ptychographic reconstructions in Fig. 3(d–f), 3(j–l) and 4(a–c). A 2D complex histogram is an extension of a normal histogram showing how many data points of a complex field lie within a certain range of real and imaginary parts. For the approximately binary test samples used in this study, ideally, the complex histograms consist of only two $\delta$-function peaks corresponding to the transmissive and opaque areas. In reality, the two $\delta$-function peaks are broadened due to limited SNR and spatial resolution. The quality of the ptychographic reconstructions can thus be assessed by examining the degree of broadening of these peaks, where reconstructions with higher quality have narrower peaks.

We first evaluate the quality of the ptychographic reconstructions in Fig. 3. In the complex histograms shown in Fig. S2, the two parts of the sample (free space and the carbon bars) are indicated by the red and yellow circles respectively. The complex histograms for OAM-HHG images (d–f) have narrower peaks than those for Gaussian-HHG images (a–c), which indicates that OAM-HHG images have better quality. The reconstruction in (c) (Gaussian-HHG illuminations on a 3-um-pitch structure) failed, thus not showing the double-peak feature.

We then evaluate the quality of the ptychographic reconstructions in Fig. 4. As shown in Fig. S3(a–c), the ptychographic reconstructions are shown in the complex representation with amplitude and phase indicated by brightness and hue. Visually, the image from OAM-HHG illumination in a has the best quality in terms of a sharp transition from free space area to thin carbon bar area and smoothness within free space or carbon bar areas. The complex histograms in (d–f) confirmed this: the primary peaks (indicated by the red and yellow circles) in the complex histogram from OAM-HHG illuminations (as shown in d) are the narrowest.

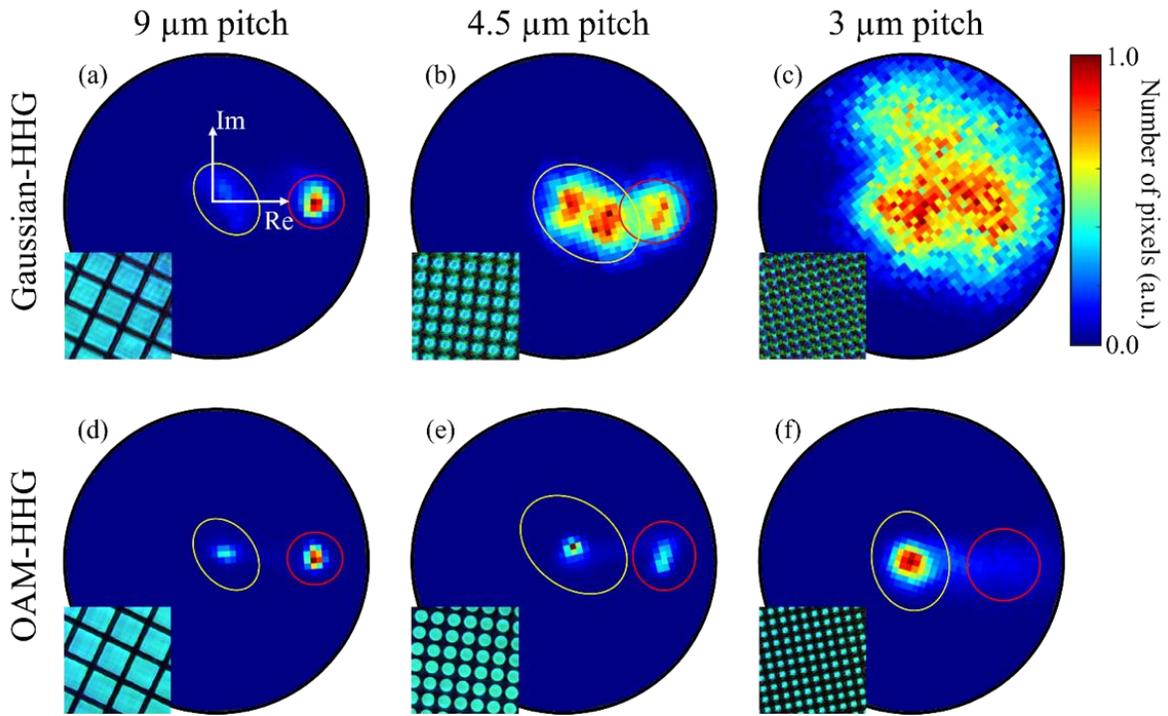

**Figure S2. Quality assessment of ptychographic reconstructions in Fig. 3 using complex histogram analysis.** (**a–c**) Complex histograms of ptychographic reconstructions of 9 μm, 4.5 μm and 3 μm pitch periodic structures using Gaussian-HHG illuminations. The ptychographic reconstructions are shown in each bottom left corner and correspond to Fig. 3(d–f). (**d–f**) Complex histograms of ptychographic reconstructions of 9 μm, 4.5 μm and 3 μm pitch periodic structures using OAM-HHG illuminations. The ptychographic reconstructions are shown in each bottom left corner and correspond to Fig. 3(j–l). These complex histograms consist of two primary peaks (except panel **c** because the reconstruction failed), which correspond to the open space area (indicated by the red circles) and the thin carbon bar area (indicated by the yellow circles). The complex histograms from OAM-HHG illuminations (the bottom row) have narrower primary peaks than those from Gaussian-HHG illuminations (the top row), which shows superior image quality for OAM-HHG reconstructions. The 'Re' and 'Im' axes in (**a**) show the complex coordinate.

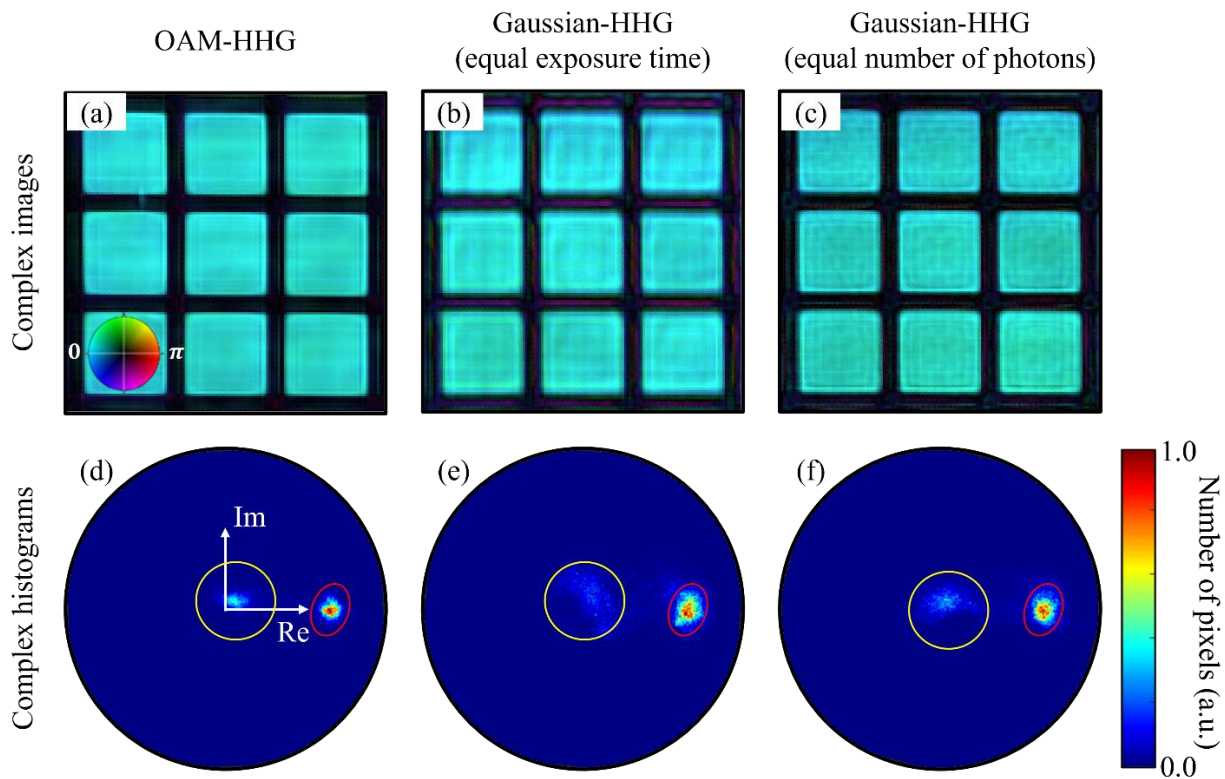

**Figure S3. Quality assessment of ptychographic reconstructions in Fig. 4 using complex histogram analysis.** (**a–c**) Complex representations of ptychographic reconstructions of 2D square periodic structures with 9 μm period under three different experimental conditions: (**a**) an OAM-HHG illumination, (**b**) a Gaussian-HHG illumination with equal exposure time, and (**c**) a Gaussian-HHG illumination with equal number of photons. The amplitude and phase of these images are presented in brightness and hue, respectively. The color wheel is shown in the bottom left corner of panel a. (**d–f**) Complex histograms of ptychographic reconstructions are shown in (**a–c**). These complex histograms all consist of two primary peaks, which correspond to the open space area, indicated by the red circles, and the thin carbon bar area, indicated by the yellow circles. The complex histogram in panel d has the narrowest primary peaks, which indicates its superior image quality provided by the intrinsic advantages of OAM-HHG illumination.

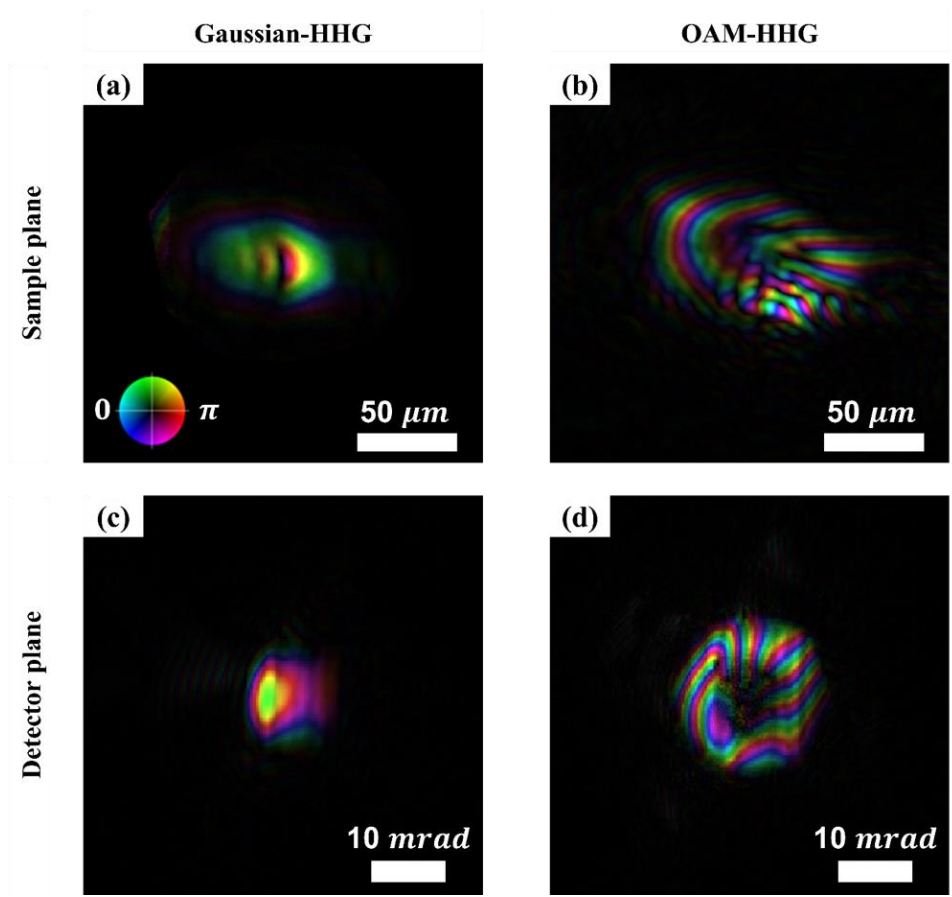

**Figure S4. Complex representations of the ptychography reconstructed Gaussian-HHG and OAM-HHG beams in the sample plane (a–b) and in the detector plane (c–d).** The amplitude and phase of the beams are shown in brightness and hue, respectively. The scale bars in (**a–b**) indicate beam size in the sample plane, and those in (**c–d**) indicate beam divergence angle in the detector plane. The OAM-HHG beam in the detector plane in (**d**) shows a characteristic donut intensity profile, while the OAM-HHG beam in the sample plane does not show a donut intensity profile due to aberrations introduced by the focusing optics.